\documentclass{eptcs}
\providecommand{\event}{QAPL 2010} 
\providecommand{\firstpage}{1}
\providecommand{\volume}{}
\providecommand{\anno}{2010}
\providecommand{\eid}{}

\def\titlerunning{Granularity-based Interface Between RTC and TA}
\def\authorrunning{K. Altisen \& Y. Liu \& M. Moy}

\usepackage{xspace}
\newcommand{\MTA}[0]{M-TA\xspace}
\newcommand{\model}[0]{\mathcal{P}}

\usepackage[utf8]{inputenc}
\DeclareUnicodeCharacter{00A0}{~}
\usepackage[T1]{fontenc}
\usepackage{paralist}

\usepackage{float}
\usepackage{placeins}
\restylefloat{figure}
\restylefloat{table}

\usepackage{graphicx}

\usepackage{latexsym,amssymb}
\usepackage{amsmath}

\usepackage{xcolor}
\usepackage{tikz}
\usetikzlibrary{positioning}
\usetikzlibrary{shapes.geometric}
\usetikzlibrary{arrows}
\usetikzlibrary{calc}
\tikzstyle{TAstate}=[circle, draw, draw=blue!70!black,
  label distance=0mm, minimum size=1cm, node distance=3cm,
  very thick,
  text justified, font=\footnotesize
]

\tikzstyle{PMCstate}=[double,circle, draw, draw=blue!70!black,
  label distance=0mm, minimum size=1cm, 
  line width=.7mm,
  text justified, font=\footnotesize
]

\tikzstyle{TAtrans}=[
  ->,draw,very thick,font=\footnotesize, 
  draw=blue!70!black, text justified
]

\tikzstyle{initial}=[
  pin={[pin distance=5mm, pin edge={<-,blue!70!black,very thick}]above:{}}
]

\tikzstyle{mode}=[dashed, draw, draw=blue!70!black,
  label distance=0mm, minimum size=.7cm, 
  very thick,
  text justified
]

\newtheorem{lemma}{Lemma}
\newtheorem{proofob}{Proof Obligation}

\newtheorem{definition}{Definition}

\newcommand{\pourreftex}[1]{}

\title{Performance Evaluation of Components Using a Granularity-based
  Interface Between Real-Time Calculus and Timed Automata}

\author{Karine Altisen \quad Yanhong Liu \quad Matthieu Moy
  \institute{Verimag, Grenoble, France.} 
  \email{\{karine.altisen, yanhong.liu, matthieu.moy\}@imag.fr}
}

\begin{document}

\maketitle

\begin{abstract}
  To analyze complex and heterogeneous real-time embedded systems,
  recent works have proposed interface techniques between {\em
    real-time calculus} (RTC) and {\em timed automata} (TA), in order
  to take advantage of the strengths of each technique for analyzing
  various components. But the time to analyze a state-based component
  modeled by TA may be prohibitively high, due to the {\em state space
    explosion} problem. In this paper, we propose a framework of
  granularity-based interfacing to speed up the analysis of a TA
  modeled component. First, we abstract fine models to work with
  event streams at coarse granularity. We perform analysis of the
  component at multiple coarse granularities and then based on RTC
  theory, we derive lower and upper bounds on arrival patterns of
  the fine output streams using the causality closure algorithm of
  \cite{causality-TACAS10}.  Our framework can help to achieve
  tradeoffs between precision and analysis time.
\end{abstract}

\section{Introduction}
\label{sec:introduction}

  




Modern real-time embedded systems are increasingly complex and
heterogeneous. They may be composed of various subsystems, and it is a
general practice that some of the subsystems may be power-managed
\cite{BBL04}, in order to reduce energy consumption and to extend the
system life time. Such a subsystem may have multiple running modes. A
mode with lower power consumption also implies lower performance
levels. Due to real-time requirements, it is thus critical to
analyze the system timing performance, 
 but the complexity of 
this  analysis is challenging, especially when it is
scaled to large and heterogeneous systems.


Compositional analysis techniques have been presented as a way of
tackling the complexity of accurate performance evaluation of large
real-time embedded systems. Examples include SymTA/S (Symbolic
Timing Analysis for Systems) \cite{symta05} and modular performance
analysis with real-time calculus (RTC) \cite{CKT03}. Various models
have also been proposed to specify and analyze
heterogeneous components \cite{PWTHSHREG07}. Each analysis
techniques have their own particular strengths and weaknesses. For
example, SymTA/S or RTC based analysis can provide hard lower/upper
bounds for the best-/worst-case performance of a system, and has the
advantage of short analysis time. But typically, they are not able to
model complex interactions and state-dependent behavior and can only
give very pessimistic results for such systems. On the other hand,
state-based techniques, e.g. timed automata (TA) \cite{Alur94atheory},
construct a model that is more accurate, and can determine exact
best-/worst-case results. But they face the {\em state
  explosion} problem, leading to prohibitively high analysis time
and memory usage
even for a system with reasonable size.


Efforts have been paid to couple different 
 approaches \cite{LSP08,
  KHET2007a, SSE07, verimag-TR-2009-14}, e.g. to combine the
functional RTC-based analysis with state-based models. 
The most
general ones are based on interfacing RTC with another existing
formalism:
in \cite{PCTT07}, an interfacing technique is proposed to
compose RTC-based techniques with state-based analysis methods 
using  {\em
  event count automata} \cite{CPT05}. A tool called CATS \cite{cats}
is just at the beginning of its development. It allows modeling a
component using TA within a system described by RTC. A variant of the
approach is described in~\cite{LPT09}, which restricts to convex and
concave curves, but potentially uses less clock and may therefore
scale better.


In this paper, we follow the line of recent development in interfacing
between RTC and state-based models done in CATS, and focus on speeding
up the analysis for a power-managed component (PMC)
modeled by TA. We adapt the framework to analyze event streams at
different granularities. This implies the ability to design the
component to be analyzed at coarser granularities: the translation of
the component from one granularity to another has to be made
automatic. To do so, we focus on a class of timed automata of interest
that we call \MTA,
on which the translation is possible. \MTA
model power-managed components, characterized with different \emph{modes} of
computation.  We then define a schema of translation at coarser
granularity. The schema is formally proven and the whole approach is
validated with experimentation: we show that our abstracted model
scales far better than the fine-granularity one, with a reasonable
loss of precision.

{\noindent\bf Organization of the paper:} in the next
section~\ref{sec:interf-TA-RTC}, we detail the implementation of the
existing interfacing techniques between RTC and TA. In
section~\ref{sec:framework}, we give an overview of the framework for
granularity-based interfacing and section~\ref{sec:changing-granularity}
formally details the granularity change. In
section~\ref{sec:power-manag-comp}, we discuss our targeted PMC, and
its fine and coarse TA models; and in section~\ref{sec:exp} we present
some experimental validations. Finally, we make a conclusion and
present future work in section~\ref{sec:conclusion}.

\section{Interfacing Timed Automata and Real-Time Calculus}
\label{sec:interf-TA-RTC}

\pourreftex{
\subsection{Real-Time Calculus (RTC)}
}
\paragraph{Real-Time Calculus (RTC).}
\label{sec:real-time-calculus}


The \emph{Real-Time Calculus (RTC)} \cite{TCN99} is a framework to model
and analyze heterogeneous system in a compositional manner. It relies
on the modeling of timing properties of event streams and available
resources with curves called \emph{arrival curves} and \emph{service
  curves}.  A component can be described with curves for its input
stream and available resources and some other curves for the
outputs. For already-modeled components, RTC gives exact bounds on the
output stream of a component as a function of its input stream.  This
result can then be used as input for the next component.


An \emph{arrival curve} is an abstraction to represent the set of
event streams that can be input to (resp. output from) a component; it
is expressed as a pair of curves $\xi=(\xi^L, \xi^U)$. For $k\geq 0$,
$\xi^L(k)$ and $\xi^U(k)$ respectively provide for any potential
stream the lower and upper bounds on the length of the time interval
during which \emph{any} $k$ consecutive events can arrive.  Let $t_i$
denote the arrival time of the $i$-th event; $t_i$ may be real (we use
continuous time) but the number of events that occurs at $t_i$ is
discrete (it is represented by a natural number); we have
$\xi^L(k)\leq t_{i+k}-t_i \leq \xi^U(k)$ for all $i\geq 0$ and $k\geq
0$. 



Similarly, the processing capacity of a component is specified by a
{\em service curve} $\psi=(\psi^L, \psi^U)$. The length of the time to
process any $k$ consecutive events for any potential stream is at
least $\psi^L(k)$ and at most $\psi^U(k)$.

Notice that, in the RTC theory, an arrival curve $\alpha=(\alpha^L,
\alpha^U)$ (resp. service curve $\beta$) is usually expressed in terms
of {\em numbers of events} per time interval. In this paper, we express
the arrival curves ($\xi$) and service curves ($\psi$) in terms of
{\em length of time interval}, in order to explain better our
work. Actually, $\xi$ is a {\em pseudo-inverse} of $\alpha$,
satisfying $\xi^U(k) = \min_{\Delta\geq 0}\{\Delta|
\alpha^L(\Delta)\geq k\}$ and $\xi^L(k) = \max_{\Delta\geq 0}\{\Delta|
\alpha^U(\Delta)\leq k\}$ (same for $\beta$ and $\psi$). 


\pourreftex{
\subsection{Timed Automata (TA)}
}
\paragraph{Timed Automata (TA).}
\label{sec:timed-automata}


A timed automaton \cite{Alur94atheory} 
is a finite-state machine extended with {\em clocks}. A clock
measures the time elapsed since its last reset and all clocks
increase at the same rate.  Figure~\ref{fig:observer}(b) shows an
example. 
States are labelled by \emph{invariants} and transitions by
\emph{guards and clock resets} (e.g. $\mathbf{t}\leftarrow
0$). Invariants and guards are conjunctions of lower and upper bounds
on clocks and differences between clocks. The automaton can only let
time elapse in a state whenever the invariant evaluates to true.  A
transition may be triggered when the guard evaluates to true and some
clocks are reset.  Timed automata may be synchronized via binary
rendez-vous: for example, the self-loop transition around
\texttt{Counting} sends the signal \textbf{produce?} This means that
the transition cannot be fired unless the automaton receives at the
same instant the signal \textbf{produce!}  from another one.
Furthermore, we use UPPAAL TA syntax and extensions \cite{cora}, such
as
integer counters (e.g. $cost$).



\subsection*{Interfacing TA and RTC}
\label{sec:interfacing-ta-rtc}







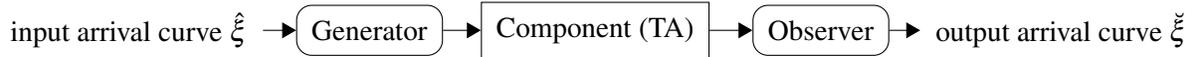
\begin{figure}[htbp]
  \centering
  \begin{tikzpicture}[inner sep=0.2cm, rounded corners=0.2cm]
    \path (-0.2,0) node (input) 
    {input arrival curve $\hat{\xi}$};
    \path (3,0) node (gen) [rectangle, draw] {Generator};
    \path (6,0) node (ta) [rectangle,  rounded corners=0cm,
    draw
    ] {Component (TA)};
    \path (9,0) node (obs) [rectangle, draw] {Observer};
    \path (12.2,0) node (output) 
    {output arrival curve $\breve{\xi}$};
    \draw [-triangle 60] (input) -- (gen);
    \draw [-triangle 60] (gen) -- (ta);
    \draw [-triangle 60] (ta) -- (obs);
    \draw [-triangle 60] (obs) -- (output);
  \end{tikzpicture}
  \vspace{-.5cm}
  \caption{Interfacing RTC and TA}
  \label{fig:cats}
\end{figure}

In the following, we show the techniques from CATS \cite{cats} for
interfacing TA and RTC. The motivation comes from the fact that some
components may not be accurately modeled with RTC tools. The idea is
then to use TA for the component model, but to be nevertheless able to
consider arrival and service curves to characterize inputs and outputs
of the system. This implies the use of adapters, as shown in
Figure~\ref{fig:cats}, connecting RTC curves (which describe a set of
event streams) and the TA (which inputs or outputs a single event
stream). From RTC curves to TA, we use a {\em generator} that inputs
events into the component such that the event stream satisfies the
input curve $\hat{\xi}$. From TA to RTC, the component outputs events
that feed an \emph{observer} which measures the smallest and largest
time interval between events, to compute the output curve
$\breve{\xi}$. Both generators and observers are timed automata; they
are composed with the timed automaton \emph{component} and the tool
UPPAAL CORA \cite{cora} is used to verify the whole.
Notice that this framework is convenient whatever be the computational
model used for the component (see e.g. ac2lus~\cite{verimag-TR-2009-14}).


In all this work, RTC curves are assumed to be given by a
finite set of $N$ points, namely, $(\rho^L,\rho^U)$ are defined by the
points $(\rho^L(i),\rho^U(i))$, for $i\in[1:N]$.

\paragraph*{Generator.}
\label{sec=generator}



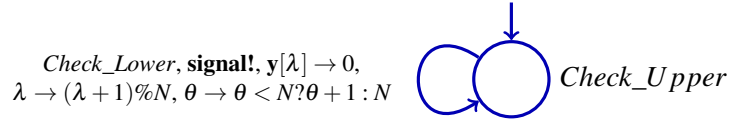
\begin{figure}[htbp]
  \centering
  \begin{tikzpicture}[]
    \node[TAstate,initial] (q) [
    label={right:$Check\_Upper$}
    ] {}; 

    \path[TAtrans] (q)  .. controls ++(-1.5,1) and ++(-1.5,-1) .. (q)
    node[pos=.5, left]
    {
      \begin{tabular}{c}
        $Check\_Lower$, \textbf{signal!}, 
        $\mathbf{y}[\lambda]\rightarrow0$, \\
        $\lambda\rightarrow(\lambda+1)\%N$,
        $\theta\rightarrow\theta<N?\theta+1:N$
      \end{tabular}
    };
  \end{tikzpicture}
  \vspace{-.5cm}
  \caption{$Generator(\rho^L,\rho^U,\text{\bf signal})$: model of
      a generator from RTC to TA.}
  \label{fig:generator}
\end{figure}

The goal for the generator is, given an input arrival (resp. service)
curve, to be able to generate \emph{any} event stream that satisfies this
curve, and only these ones.
Figure~\ref{fig:generator} shows the TA model
$Generator(\rho^L,\rho^U,\text{\bf signal})$ which non
deterministically generates a sequence of signals called
\textbf{signal!} that satisfies $(\rho^L,\rho^U)$.

The main idea is to reset a clock whenever an event is emitted. As we
need to check any $N$ successive events, we need to memorize only $N$
clocks, that we declare in a circular clock array $\mathbf{y}$ of size $N$.
$\mathbf{y}[k\%N]$ represents the time elapsed since the event $k$ has
occurred. We note $\lambda$ the index of the next event to be generated:
then we have to check that the $N$ events before $\lambda$ comply with
the curve. The index of those events is given by the 
function
$getIdx(i) \rightarrow(\lambda-i+N)\%N \quad (1 \leq i \leq N)$.
Note that at the beginning, less than $N$ events have been emitted: we
introduce a bounded counter $\theta$, from $0$ to $N$ that represents
the actual number of events to be considered.  The constraints that
satisfy the curves are then expressed as:
\begin{eqnarray*}
  (Check\_Upper) & \mathbf{y}[getIdx(i)]\leq \rho^U(i), \forall
  i\in[1:\theta] \\ 
  (Check\_Lower) & \mathbf{y}[getIdx(i)]\geq \rho^L(i), \forall
  i\in[1:\theta] 
\end{eqnarray*}
$(Check\_Upper)$ expresses an invariant on how much time can
elapse (i.e. when an event \emph{must} be generated), whereas
$(Check\_Lower)$ qualifies the date from which a new event \emph{may}
be emitted, checked before emitting a new signal.

\paragraph*{Observer.}

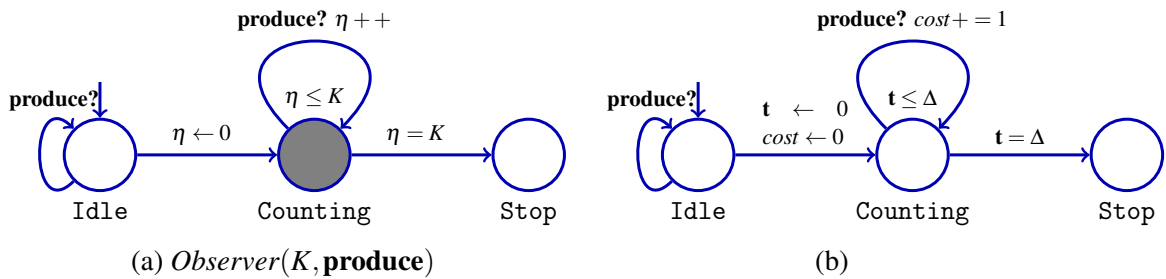
\begin{figure}[htbp]
  \vspace{-.2cm}
  \begin{minipage}[t]{0.5\linewidth}
    \centering
    \scalebox{.95}{
  \begin{tikzpicture}
    \node[TAstate,initial] (idle) [
    label={below:\texttt{Idle}},
    ] {}; 
    
    \node[TAstate, fill=gray] (busy) [
    right of=idle,
    label=below:{\texttt{Counting}},
    label={ 
      above:
      \footnotesize$\eta \leq K$ 
    } 
    ] {};

    \node[TAstate] (stop) [
    right of=busy,
    label=below:{\texttt{Stop}}
    ] {}; 
  
    \path[TAtrans] (idle) -- (busy) 
    node[midway, above, text width=1cm] 
    {$\eta \leftarrow 0$};

    \path[TAtrans] (busy) -- (stop) 
    node[midway, above, text width=1cm] 
    {$\eta=K$};
    
    \path[TAtrans] (busy) .. controls ++(-2,2) and ++(2,2) .. (busy) 
    node[midway, above]
    {
      \textbf{produce?} $\eta++$};
    
    \path[TAtrans] (idle) .. controls ++(-1,-1) and ++(-1,1) .. (idle) 
    node[pos=.8,above, text width=1.2cm] 
    {\textbf{produce?}};
    
  \end{tikzpicture}
  }

  (a) $Observer(K,\text{\bf produce})$
  \end{minipage}
  \begin{minipage}[t]{0.4\linewidth}
    \centering
    \scalebox{.95}{
  \begin{tikzpicture}
    \node[TAstate,initial] (idle) [
    label={below:\texttt{Idle}},
    ] {}; 
    
    \node[TAstate] (busy) [
    right of=idle,
    label=below:{\texttt{Counting}},
    label={ 
      above:\footnotesize
      $\mathbf{t} \leq \Delta$
    } 
    ] {};

    \node[TAstate] (stop) [
    right of=busy,
    label=below:{\texttt{Stop}}
    ] {}; 
  
    \path[TAtrans] (idle) -- (busy) 
    node[midway, above, text width=1.2cm] 
    {$\mathbf{t} \leftarrow 0$ $cost \leftarrow 0$};

    \path[TAtrans] (busy) -- (stop) 
    node[midway, above]
    {$\mathbf{t}=\Delta$};
    
    \path[TAtrans] (busy) .. controls ++(-2,2) and ++(2,2) .. (busy) 
    node[midway, above]
    {
      \textbf{produce?} $cost += 1$};
    
    \path[TAtrans] (idle) .. controls ++(-1,-1) and ++(-1,1) .. (idle) 
    node[pos=.8,above, text width=1.2cm] 
    {\textbf{produce?}};
    
  \end{tikzpicture}
  }

  (b)
  \end{minipage}
  \vspace{-.1cm}
  \caption{(a) Model of the observer and (b) its varied model.}
  \label{fig:observer}
\end{figure}

To compute the output curve $(\rho^L,\rho^U)$, we use a model checking
tool with cost optimal reachability analysis~\cite{cora}. An
observer is used to capture the arrival patterns of an
event stream: Figure~\ref{fig:observer}(a) computes lower and upper bound
for a stream of \textbf{produce!} events and for a window of $K$ events,
namely $(\rho^L(K),\rho^U(K))$.

The observer non-deterministically transits from
the initial state \texttt{Idle} to \texttt{Counting}: this decides the
beginning of the window to be observed. The counter $\eta$ records the
number of \textbf{produce!} since the entry in \texttt{Counting}. When
reaching the state \texttt{Stop}, $K$ \textbf{produce!} events have
been emitted.

The cost for an execution corresponds to the time spent in the
\texttt{Counting} state (no cost on transition, cost rate of 1 for
\texttt{Counting} -- marked as a grey state in the figure, cost rate
of 0 for the other states). Therefore, the cost when reaching the
state \texttt{Stop} is equal to the time for emitting $K$ events. With
a verification engine able to compute the minimum and maximum cost for
reaching \texttt{Stop}, this provides $\rho^L(K)$,
since the $K$ consecutive events were chosen
non-deterministically. The computation has to be launched $N$ times
(for $K = 1$ to $N$) to obtain all the points of the curves.

Since the tool we use can not compute a maximum, we use a variant
model of the observer shown in Figure~\ref{fig:observer} (b) for
computing the maximum and obtaining $\rho^U(K)$. The principle of
the timed automata is the same as the former but
it measures the number of events (counter $cost$) that can be emitted
during an interval of length $\Delta$ (clock $\mathbf{t}$). When
minimizing the cost, this provides $\alpha^U(\Delta)$.  Then, $\rho^U$
is computed as the pseudo-inverse of $\alpha^U$.


\paragraph*{How to obtain output arrival curves?}


We show here how to instantiate the timed automata described above to
compute an output arrival curve $\breve{\xi}$ given
the input arrival curve $\hat{\xi}$,
the input service curve $\psi$
and a component modeled as a timed automaton.
The input event stream is labeled by signals \textbf{req} and is
given by $Generator(\hat{\xi}^L,\hat{\xi}^U, \text{\bf req})$. The
input service stream is labelled by signals \textbf{serv} and is
generated by $Generator(\psi^L,\psi^U, \text{\bf serv})$.  We assume
that the component inputs \textbf{req} and \textbf{serv} signals and
emits \textbf{produce} signals. The output event stream is then
analyzed by $Observer(K,\text{\bf produce})$.
Those four automata are synchronized and a verification engine is
used, to obtain all the points
$(\breve{\xi}^L(K),\breve{\xi}^U(K))$, for $K = 1$ to $N$.



Note that the whole model involves two clock arrays of size $N$ (length
of the curve); it also involves two $N$-counters per generator plus
one $N$-counter for the observer. The size of the model, hence, heavily
depends on the number of events to be considered.





\section{The Granularity-based Interfacing Framework}
\label{sec:framework}


\paragraph{Motivations.}

When model-checking non-trivial systems of TA, one quickly faces the
well-known state-explosion problem. Generally speaking, this
mainly comes from two sources: clocks and counters. 
Counters are handled as discrete states 
in the proof engine and clocks
involve the dimensions of the polyhedra (zones) to be computed.
As stated above, the CATS approach proposes to model-check (with cost
optimization) a model where the numbers of clocks and the order of magnitude of
counters are heavily linked with the number of events to be considered
and so is the cost to compute the results.  Applied to non-trivial
components the approach may thus fail to provide any result.

By grouping the events, and refraining from looking at them
individually, we can reduce 
the amount of events analysis and thus both the size of counters and
the number of clocks: we can hence get dramatic improvements on the
performance. 
%
We talk about \emph{fine events} to designate the real events of the
system and we group them into \emph{coarse events} which intuitively
represent a packet of $g$ fine events, where $g$ is called the
granularity of the abstraction. Modeling the system to work with
coarse events instead of fine events divides the length of the arrival
curve $\hat{\xi}$ and the number of events $N$ to store in buffers by
$g$. By changing the value of $g$, one can trade performance for
accuracy.

\label{sec:granu-approach}

\paragraph{Framework.}

We propose a formal framework of granularity-based
interfacing between RTC and TA performance models. The generators for
service and arrival curves, the component model and the observer for
the output arrival curve deal all with coarse events, which speeds up
the analysis. 

\begin{figure}[htbp]
  \centering
  \begin{tikzpicture}
    \tikzstyle{textbox}=[text width=3cm,rectangle,draw,text centered]
    \tikzstyle{legend}=[above]
    \node (input)  at (0,0) [textbox] {
      Model with modes};
    \node (start)  [above=5mm of input]{};
    \node (fine)   at (5.5,0) [textbox] {Fine-grain TA};
    \node (coarse) at (-5.5,0) [textbox] {Coarse-grain TA};
    \node (inputcurve) at (-8.5,-2) {$\hat{\xi}$};
    \node (analysis) at (-5.5,-2) [textbox, rounded corners=0.5cm] {%
      Analysis at multiple granularities $g_1, \ldots, g_m$};
    \node (combination) at (0, -2) [textbox] {Combination};
    \node (result) at (4, -2) [anchor=west] {$\breve{\xi}$ (output curve)};

    \path[->]
    (start)    edge (input)
    (input)    edge node[legend] {\small Semantics} node[below] {\small (a)} (fine)
    (input)    edge node[legend] {\small Abstraction} node[below] {\small (b)} (coarse)
    (coarse)   edge (analysis)
    (inputcurve) edge (analysis)
    (analysis) edge node[legend] {\small
      $\breve{\xi}_{g_1}...\breve{\xi}_{g_m}$} node[below] {\small (c)} (combination)
    (combination) edge (result);
  \end{tikzpicture}
\caption{The framework of granularity-based interfacing.}
\label{fig:scheme}
\end{figure}
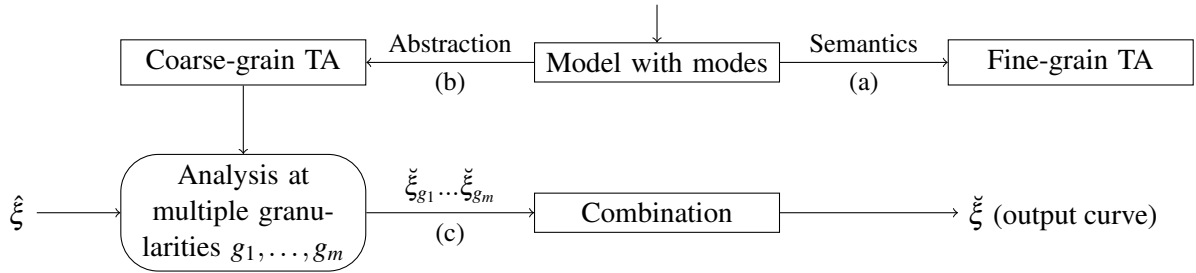

As we change the granularity, all TA models involved in the
framework have to be adapted to deal with the coarse events. For the
generators and the observers, this is quite straightforward, but
abstracting an arbitrary TA component to a coarse-granularity one
would be hard, if at all possible.  We focus on a particular class of
timed automata that we call \MTA (for ``Mode-based Timed-Automata'') for
which we propose an automatic translation scheme from fine event \MTA
to coarse event \MTA. For simplicity of the notations, we define the
class of \MTA with an abstract syntax, the semantics of an \MTA being 
given by the corresponding TA. 
This is illustrated by (a) and (b) in Figure~\ref{fig:scheme}.
The correctness of the approach relies on the fact that the
coarse-grain translation of an \MTA is an accurate abstraction, in the
sense that the lower and upper coarse output curves, analyzed from the
coarse model, always provide lower and upper bounds on the lower and
upper fine curves, which would be obtained from the original fine
model. This is proved once and for all for the translation scheme.

When analyzing a coarse model, we get a pair of coarse output
arrival curves that already provides lower and upper bounds on the
arrival patterns of the output stream at fine granularity, by applying
a simple rescaling. But 
it is possible to derive tighter fine output curve, running the
analysis at different granularity levels $g_1...g_m$.  The resulting
coarse output curves $\breve{\xi}_{g_1}...\breve{\xi}_{g_m}$ are then
combined (see Figure~\ref{fig:scheme}.(c)) using 
the causality closure \cite{causality-TACAS10} property of RTC
curves. This results in a
tighter output arrival curve ($\tilde{\xi}$) at the fine granularity
which is tighter than the naive combination of the curves but still
equivalent to it.

To the best of our knowledge, no existing work deal with
a granularity-based scheme with formal validation on the
abstraction. The proposed framework complements the recent works on
interfacing between RTC and TA models.

\section{Changing the Granularity}
\label{sec:changing-granularity}

\paragraph{Definitions and Notations.}

We denote a specific stream at the fine granularity as
$\tau = (t_0)t_1t_2...$ and a specific coarse stream to be $\top =
(T_0)T_1T_2...$. $t_i$ (resp. $T_i$) denotes the arrival time of the
$i$-th fine (resp. coarse) event $\varepsilon_i$ (resp.
$\mathcal{E}_i$) for $i\geq 1$. $t_0=T_0=0$ denotes the origin of
time. $\hat{\top}$ and $\hat{\tau}$ denote the input, while
$\breve{\top}$ and $\breve{\tau}$ denotes the output stream.

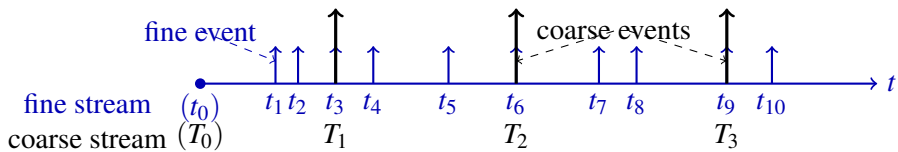
\begin{figure}[htbp]
  \centering

  \begin{tikzpicture}[draw=blue!70!black,color=blue!70!black]
    \path (1,0) coordinate (t0) node[below] {$(t_0)$};
    \path (2,0) coordinate (t1);
    \path (2.3,0) coordinate (t2);
    \path (2.8,0) coordinate (t3);
    \path (3.3,0) coordinate (t4);
    \path (4.3,0) coordinate (t5);
    \path (5.2,0) coordinate (t6);
    \path (6.3,0) coordinate (t7);
    \path (6.8,0) coordinate (t8);
    \path (8,0) coordinate (t9);
    \path (8.6,0) coordinate (t10);
    \path (10,0) coordinate (etc);
    \fill (t0) circle (2pt);
    \draw[->,thick] (t0) -- (etc.east) node[right] {$t$};
    \foreach \i in {1,...,10} { \draw (t\i) node[below] {$t_{\i}$}; };
    \foreach \i in {1,...,10} { \draw[->,thick] (t\i) -- ($(t\i)+(0,.5)$); };
    \path ($(t0)+(0,-.7)$) node[color=black] {$(T_0)$};
    \path ($(t3)+(0,-.7)$) node[color=black] {$T_1$};
    \path ($(t6)+(0,-.7)$) node[color=black] {$T_2$};
    \path ($(t9)+(0,-.7)$) node[color=black] {$T_3$};
    \draw[->,very thick,draw=black] (t3) -- ($(t3)+(0,1)$);
    \draw[->,very thick,draw=black] (t6) -- ($(t6)+(0,1)$);
    \draw[->,very thick,draw=black] (t9) -- ($(t9)+(0,1)$);
    \path (-.5,0) coordinate (label) node[below] {fine stream} 
    node[below=.5,color=black] {coarse stream};

    \draw[->,dashed] (1,0.7) -- ($(t1)+(0,.3)$) node[pos=0] {fine event};
    \draw[->,dashed,draw=black] (6.5,0.7) -- ($(t6)+(0,.3)$) 
    node[pos=0,color=black] {coarse events};
    \draw[->,dashed,color=black] (6.7,0.7) -- ($(t9)+(0,.3)$);
 \end{tikzpicture}
  \caption{Illustration of a fine and coarse stream with $g=3$.}
\label{fig:destream}
\end{figure}

As illustrated in Figure~\ref{fig:destream}, we can abstract a fine
stream to a coarse one at some granularity $g$, by regarding $g$
consecutive fine events as a coarse one. A fine stream $\tau$ is a
\emph{refinement} of a coarse one $\top$ if $T_i$ is equal to $t_{gi}$
for $i\geq 0$, i.e. if it is sampling the fine stream every $g$
events. A coarse stream $\top$ at granularity $g$ is an
\emph{abstraction} of a fine stream if $t_{gi}=T_i$ for $i\geq 0$, the
other $t_k$ being chosen arbitrarily, with $t_{k+1} \geq t_k$.  It is
easy to see that a coarse stream $\top$ can be refined to a fine
stream $\tau$ if and only if $\tau$ can be abstracted to $\top$.

\label{sec:abstractions}

\paragraph{How to obtain coarse output arrival curves?}
We now show how to use the interfacing between RTC and TA
(Section~\ref{sec:interf-TA-RTC}) to compute output arrival curves at
coarser granularity.  The input arrival (resp. service) curve
$\hat{\xi}$ (resp. $\psi$) is given for fine events.
%
%
\label{sec:sampling}
To obtain the input arrival (resp. service) curve for coarse events at
granularity $g$, $\hat{\xi_g}$ (resp. $\psi_g$), we use a
\emph{sampler}. From $\hat{\xi}$ it produces $\hat{\xi_g}$ such that
$\hat{\xi}_g(k) = \hat{\xi}(gk)$ for $k\geq 0$ (the definition of
$\psi_g$ is exactly the same).  Indeed, it is easy to see that the
arrival (resp. service) pattern for the coarse input streams are lower
and upper bounded by $\hat{\xi}(gk)$ (resp. $\psi(gk)$) for all $k\geq
0$. Notice that this abstraction from fine input \emph{streams}
$\hat{\tau}$ to coarse ones $\hat{\top}$ at granularity $g$ implies to
lose information about every fine events but the ones at $gi$,
$\varepsilon_{gi}$.



To compute a coarse output arrival curve $\breve{\xi_g}$, we then
proceed as before: a generator is used for both the arrival curve
$\hat{\xi_g}$ and the service curve $\psi_g$; and observers are used
to compute the result using model-checking with cost
optimality. Nevertheless, between them, the \emph{component} still has 
to be adapted.  Indeed, it was designed to proceed on fine events and
needs to be abstracted to work on coarse input streams.  In the
sequel, we note $\model_0$ the fine component and $\model_C$ its
coarse abstraction.  This transformation from $\model_0$ to $\model_C$
may be hard, if possible and we define a particular class of interest,
called \MTA, for which we provide an automatic transformation
scheme. This is the topic of the following section.  Notice that such
an abstraction
introduces non-determinism in the coarse component. For example,
when the triggering condition of a transition depends on a number of
events, the coarse grain component cannot know exactly the time at
which the transition is triggered.


\paragraph{Validation.}

To validate the proposed framework, one has to guarantee that the
coarse models provide \emph{accurate abstraction} of the fine
models. The proof can be made on every parts of the model but the
component, since the TA are given. For the component, we exhibit a
proof obligation that has to be guaranteed to validate the whole
framework. This proof obligation states that the coarse component
$\model_C$ should exhibit at least all the behaviors that the fine
component $\model_0$ can produce.
Formally, it should satisfy the following property.
\begin{definition}
  Let $\hat{\tau}=(\hat{t}_0)\hat{t}_1\hat{t}_2...$
  ($\breve{\tau}=(\breve{t}_0)\breve{t}_1\breve{t}_2...$) be any fine
  event stream that is an input to (the corresponding output stream
  produced from) $\model_0$.  We say that $\model_C$ is a
  \emph{correct abstraction} of $\model_0$ iff(def) there always
  exists some coarse event stream $\hat{\top}$ ($\breve{\top}$) that
  can be an input to (the corresponding output stream produced from)
  $\model_C$, such that $\hat{\top}$ ($\breve{\top}$) can be refined
  to $\hat{\tau}$ ($\breve{\tau}$).
\end{definition}
\begin{proofob}
  \label{lemma:pmc}
  Prove that $\model_C$ is a \emph{correct abstraction} of $\model_0$.
\end{proofob}

We will show later that the TA generated by our translation verify
this proof obligation. Any other way to generate coarse TA that
satisfy it could be used in the framework.

The correctness of the abstraction implies that we can derive a valid
coarse output curve $\breve{\xi}_g$, by analyzing $\model_C$.
\begin{lemma}\label{lemma:observer} 
  If Proof Obligation~\ref{lemma:pmc} is satisfied, it is guaranteed
  that the analyzed $\breve{\xi}_g^L(k)$ and $\breve{\xi}_g^U(k)$
  provide lower and upper bounds on $\breve{\xi}^L(gk)$ and
  $\breve{\xi}^U(gk)$ respectively for $k\geq 0$.  
\end{lemma}
\textit{Sketch of proof:} it follows from Proof
Obligation~\ref{lemma:pmc} that, for any fine output stream
$\breve{\tau}$ from $\model_0$, there always exists a coarse output
stream $\breve{\top}$ such that $\breve{\top}$ can be refined to
$\breve{\tau}$. It follows that the production time of the ($g\times
k$)-th fine event in $\breve{\tau}$ is equal to that of the $k$-th
coarse event in $\breve{\top}$. It is then easy to show that
$\breve{\xi}_g^L(k) \leq \breve{\xi}^L(gk)$ and $\breve{\xi}_g^U(k)
\geq \breve{\xi}^U(gk)$.
%


\paragraph{How to combine multiple coarse curves to obtain a fine one?}
\label{sec:refinement-algorithm}

The above paragraphs show how to obtain a coarse pair of output curves
at a given granularity and prove their accurateness. We now propose
to conduct \emph{multiple runs of analysis at different granularities}
$g_1, ..., g_m$, and show how to combine them to obtain a valid
fine pair of output curves.  The coarse output arrival curve at
granularity $g_i$ is noted $\breve{\xi}_{g_i}$ ($i=1,...,m$) and we
denote by $\breve{\xi}^U$ and $\breve{\xi}^L$ the optimal \emph{fine}
output arrival curves.  Due to Lemma~\ref{lemma:observer}, we have
$\breve{\xi}^U_{g_i}(k)\geq \breve{\xi}^U(kg_i)$ and
$\breve{\xi}^L_{g_i}(k)\leq \breve{\xi}^L(kg_i)$ for $k\geq 0$.

Therefore, a first approximation of $\breve{\xi}^U$ (resp.
$\breve{\xi}^L$) is to take the minimal (resp. maximal) obtained
values for different granularities: $\breve{\xi}^U_{combined}(n) =
\min_{kg_i \geq n} \{ \breve{\xi}^U_{g_i}(k) \}$.  But this curve is not
necessarily the tightest we can find.

Suppose, for example, that we obtain output curves $\breve{\xi}_2$ and
$\breve{\xi}_3$ after analyzing $\model_C $ with $g=2, 3$. 
We can get on $\breve{\xi}(1)$ some tighter bounds than the minimum of
$\breve{\xi}_2^U(1)$, $\breve{\xi}_3^U(1)$ as follows.
Let $\tau=t_0t_1t_2...$ denote the trace of the fine output stream,
($t_i$ denotes the production time of the $i$-th fine event
$\varepsilon_i$), with $t_0=0$. 
Recalling that $\breve{\xi}^L(k)$ and $\breve{\xi}^U(k)$ are defined
to be lower and upper bounds on the length of the time interval during
which any $k$ consecutive events are output, for any $i\geq 0$, we
have $\breve{\xi}^L(2) \leq t_{i+3} - t_{i+1} \leq \breve{\xi}^U(2)$ and
$\breve{\xi}^L(3) \leq t_{i+3} - t_i \leq \breve{\xi}^U(3)$. We can
thus derive constraints on time interval of length 1:
\begin{equation*}
  \label{ineq:ref1}
  \breve{\xi}^L(3) - \breve{\xi}^U(2) \leq t_{i+1}-t_{i} \leq
  \breve{\xi}^U(3) - \breve{\xi}^L(2)
\end{equation*}
Hence, a better valid bound for the time interval between any two
events is given by 
\[\left[ \max\left\{ \breve{\xi}^L(1),
    \breve{\xi}^L(3) - \breve{\xi}^U(2) \right\}, \min\left\{
    \breve{\xi}^U(1), \breve{\xi}^U(3) - \breve{\xi}^L(2) \right\}
\right]\]

Other linear constraints can be used to derive constraints in such a
flavor. But, we actually use the causality closure algorithm given
in~\cite{verimag-TR-2009-15} which refine an arbitrary pair of curves
to the optimal equivalent pair of curves (provided a slight adaptation of the
algorithm for finite curves  to work with pseudo-inverse of
the curves). This algorithm is based on the notion of deconvolution, which
is basically a generalization of the above example by taking all the
possible values instead of just $2$ and $3$ as in the example.

\section{Application to Power Managed Components}
\label{sec:power-manag-comp}
\label{sec:model}







\pourreftex{
\subsection{Power-Managed Components in Embedded Systems}
}

The systems we target to define an automatic translation from fine to
coarse models are energy-aware, or ``power-managed''. They
have different modes of operations, in which their performance and
energy consumption are defined. Most computers and embedded systems today
possess energy-saving modes; for example, CPUs can have dynamic
voltage and frequency scaling (DVFS) and sensors in a network can switch
their radio on and off.



\pourreftex{
\subsection{Requirements to Model Power-Managed Components}
}

\subsection{Models of PMC}

We consider power-managed components (PMC) as systems with
different modes of operation, each mode owning a pair of service
curves. The system can transit from a mode to another upon various
conditions. The minimal requirements to model non-trivial systems is:
\begin{compactenum}
\item[(1)] 
by receiving an explicit synchronization from another component 

\item[(2)]
after a given timeout (typically, systems go to a hibernation
  state only after spending some time in an idle state),

\item[(3)]
when the buffer fill level exceeds a certain threshold (to
  switch to a resource-intensive one when the system is overloaded),

\item[(4)]
 when the buffer fill level gets below a threshold (typically, go
  to an energy-saving state when the buffer is empty). 

\end{compactenum}
Also, we need a way to force the system to stay in a given mode for a
minimum amount of time (for example, to model a transition between two
modes that physically takes some time). This is modeled by modifying
the last two conditions to enable them only when the time spent in the
current mode is greater than some constant.



\begin{figure}[htbp]

  \centering
    \vspace{-.2cm}

  \begin{tikzpicture}
    \node[mode] (mi) {$M_i$};
    \node[mode] (ml) at +(-3cm,-1cm) {$M_l$};
    \node[mode] (mp) at +(3cm,-1cm) {$M_p$};
    \node[mode] (mk) at (mi) [left=4.5cm] {$M_k$};
    \node[mode] (mj) at (mi) [right=4.5cm] {$M_j$};

    \path[TAtrans] (mi) -- (ml)
    node[midway, sloped, below]
    {(1) $\;$ \textbf{a?}};

    \path[TAtrans] (mi) -- (mp)
    node[pos=.6, below, sloped]
    {(2) $\;\mathbf{x}=U_i$};
    
    \path[TAtrans] (mi) -- (mk)
    node[midway, above]
    {(3) $\;L_i\leq \mathbf{x} \leq U_i$ $\wedge\; q<b^L_i$};
    
    \path[TAtrans] (mi) -- (mj)
    node[midway, above, sloped]
    {(4) $\;L_i\leq \mathbf{x} \leq U_i \wedge\; q> b^U_i$};
    
  \end{tikzpicture}

  \caption{\MTA: 4 kinds of transitions between modes}
  \label{fig:desp}
\end{figure}
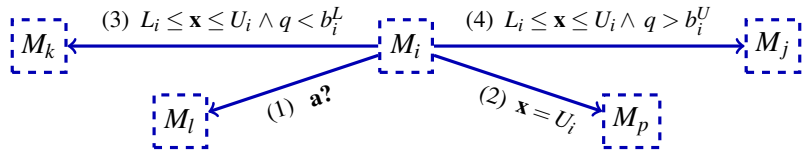

\pourreftex{
\subsection{\MTA: a Graphical Syntax to Describe PMCs}
}
\paragraph{Graphical Syntax to Describe PMC.}
From these requirements, we define the class of automata \MTA, and
give a graphical syntax for it.  The description is based on an
enumeration of modes, each mode $M_i$ being associated with a pair of
service curves $\psi_i=(\psi^L_i, \psi^U_i)$; it is restricted to how
the PMC can evolve from mode to mode.  To handle the events to be
computed, we suppose that the PMC is equipped with a buffer at its
entry. We note $q$ the counter for the buffer fill level: in each mode
$M_i$ it is constrained to be within some lower and upper bounds
$b_i^U$ and $b_i^L$.  A mode $M_i$ is also equipped with time
constants $L_i, U_i$ which represent the lower and upper time the
component can stay in the mode; we use the clock $\mathbf{x}$ to
measure this time: it is reset when entering a mode and checked
against the bounds when going out.  Figure~\ref{fig:desp} shows a
descriptive model of the PMC with all the possible mode changes
((1)..(4), using the same numbering as above).  We believe that our
work can be easily extended to other cases of mode switch.





%
%



\pourreftex{
\subsection{Semantics of \MTA{}s in Terms of TA}
}
\paragraph{Semantics of \MTA.}

Given a PMC described by a \MTA, we give its semantics in terms of the
TA it represents. Using the same convention as before, a PMC receives
events through the signal \textbf{req?} from the generator and emits
\textbf{produce!} at its output. Internally, we translate the \MTA
using two synchronized TA: the first,  called \emph{Processing
Element (PE)}, controls the mode switches 
and the second, called \emph{Service Model (SM)} models the
computation resources
for each mode. The PE and the SM are strongly synchronized: the PE
emits signals \textbf{Syn$_{ij}$!}  whenever transiting from mode
$M_i$ to mode $M_j$ and the SM changes its behavior accordingly. The
SM has the same discrete structure as the abstract syntax. It uses one
instance of the generator described in section~\ref{sec=generator} per
mode $M_i$ (i.e. per state),
$Generator(\psi_i^L,\psi_i^U,\text{\bf{serv}})$ which emits
\textbf{serv!} whenever the PMC is able to process an event. This
\textbf{serv!}  signal is then transmitted to the PE, which decrements
its backlog $q$ and emits a \textbf{produce!} (received by the
observer).

\begin{figure}[htbp]

  \begin{minipage}[c]{.4\linewidth}
    \centering
      \begin{tikzpicture}
        \node[PMCstate] (q) [
        label=above:{$[b^L,b^U]$},
        ] {$x\leq U$};
        \node (nameq) at (q) [above=1mm] {$S$}; 

        \node (equiv) at (q) [right=.5cm]
        {$\stackrel{\text{def}}{\Longleftrightarrow}$};

        \node[TAstate] (qp) at (equiv) [right=.5cm] {$x\leq U$};
        \node (nameqp) at (qp) [above=1mm] {$S$}; 
        
        \path[TAtrans] (qp) .. controls ++(1,1) and ++(1,0.2) .. (qp) 
        node[midway, right, text width=1cm] 
        {$q<b^U$ \textbf{REQ}};
        
        \path[TAtrans] (qp) .. controls ++(1,-0.2) and ++(1,-1) .. (qp) 
        node[midway, right, text width=1cm]
        {$q>b^L$ \textbf{SERV}};
      \end{tikzpicture}
    \end{minipage}
    \begin{minipage}[c]{.6\linewidth}
      \small
      \textbf{REQ} means: $\{$ \textbf{req?} $q$\texttt{++;} $\}$

      \vspace{.5em}
      \textbf{SERV} means: \texttt{if (}$q$\texttt{>0)}
      $\{$ \textbf{serv?} $q$\texttt{-{}-;} \textbf{produce!} $\}$
      \texttt{else} $\{$ \textbf{serv?} $\}$
    \end{minipage}
  \caption{Simplified notation of a state of the PE}
  \label{fig:defstate}
\end{figure}
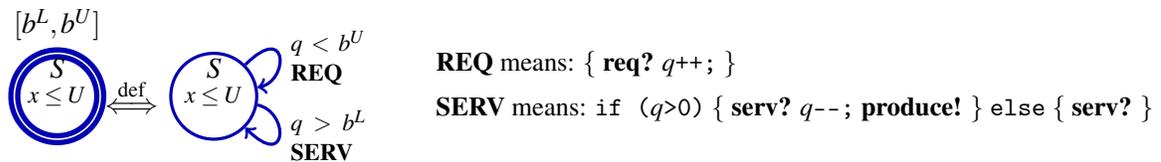

Similarly, each time the PE receives an event to be processed via the
signal \textbf{req?} (received from the generator that models the
input arrival curves) this increments the backlog $q$.
Before further detailing the implementation of the PE in terms of
timed automata, we define a shortcut syntax for a state in
Figure~\ref{fig:defstate}. The interval $[b^L, b^U]$ specified on the
state $S$ means that the system can stay in $S$ all along the buffer
fill level $q$ is within this range. The clock invariant
$\mathbf{x}\leq U$ has the usual meaning. This shortcut enables the
model to be complete with respect to the incoming events \textbf{req?}
and \textbf{serv?}.

\begin{figure}[htbp]
  \centering

  \begin{tikzpicture}[
    font=\footnotesize]
    \node[PMCstate] (Si) [
        label=above left:{$[-\infty,\infty]$},
        ] {$x\leq L_i$};
        \node (nameSi) at (Si) [above=1mm] {$S_i$}; 

    \node[PMCstate, right of=Si,node distance=4.5cm] (Si1) [
    label=above:{$[b_i^L,b_i^U]$},
    ] {$x\leq U_i$};
    \node (nameSi1) at (Si1) [above=1mm] {$S_{i1}$}; 
    
    \node[mode, right of=Si1,node distance=4.5cm] (Mj) {$M_j$};
    \node[mode, below right of=Si1,node distance=3.5cm] (Mp) {$M_p$};
    \node[mode, below left of=Si1,node distance=3cm] (Ml) {$M_l$};
    \node[mode, below of=Si1,node distance=2.7cm] (Mk) {$M_k$};
    \coordinate[below left of=Si,node distance=3cm] (Mr);
    \node[dashed, draw, draw=blue!70!black, very thick, minimum
    size=1.5cm] (Mrbis) at (Mr) {};
    \path ($(Mr)+(-.4,-.4)$) node {$M_r$};
    \path ($(Mr)+(.35,.4)$) node {$Syn_{ri}!$};

    \path[TAtrans,dashed] (Mr) |- (Si) node[pos=.7,above]
    {$x\leftarrow 0$};
    \path[TAtrans] (Si) -- (Si1) node[midway,above] {$x=L_i\wedge
      b_i^L\leq q \leq b_i^U$};
    \path[TAtrans] (Si1) -- (Mj) node[midway,above] {$q=b_i^U\;\;Syn_{ij}!$};
    \path[TAtrans] (Si1) -- (Mp) node[midway,above,sloped] 
    {$x=U_i\;\;Syn_{ip}!$};
    \path[TAtrans] (Si1) -- (Mk) node[pos=.6,right,text width=.9cm] 
    {$q=b_i^L$  $Syn_{ik}!$};
    \path[draw,very thick,draw=blue!70!black] (Si) -- ($(Si)+(2,-1)$) -- (Si1);
    \path[TAtrans] ($(Si)+(2,-1)$) -- (Ml) node[midway,right,text width=.7cm] 
    {$\mathbf{a!}\;\;Syn_{il}!$};
    \path[TAtrans] (Si) |- ($(Si)+(0,1.3)$) -| (Mj) node[pos=.3,above] 
    {$x=L_i\wedge q>b_i^U \;\; Syn_{ij}!$};
    \path[TAtrans] (Si) |- (Mk) node[pos=.7,below] 
    {$x=L_i\wedge q<b_i^L \;\; Syn_{ik}!$};

  \end{tikzpicture}
  \caption{Fine TA model of the PE}
\label{fig:finePE}
\end{figure}
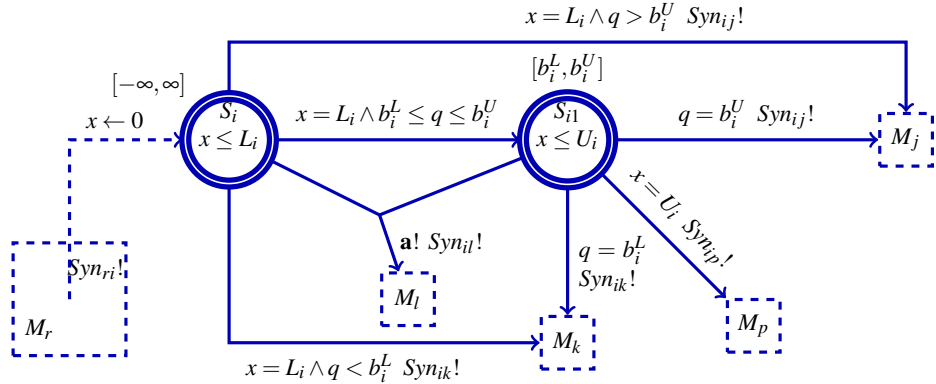

Figure~\ref{fig:finePE} shows, for the PE, the translation of the 4
kinds of transitions of the \MTA in Figure~\ref{fig:desp} into plain
TA; notice that it only expands the single mode $M_i$.
%
This mode is implemented with two states $S_i$ (initial state)
and $S_{i1}$. $S_i$ is added to ensure that the PE doesn't leave the
mode before reaching its lower timing constraint $L_i$. When
$\mathbf{x}=L_i$, it is time to leave $S_i$: either the exit condition
on $q$ ($q>b^U_i$ or $q < b^L_i$) is already satisfied and the TA
immediately switches to the corresponding mode ($M_j$ or $M_k$) or it
transits to the state $S_{i1}$. It can stay in $S_{i1}$ while
$q\in[b^L_i, b^U_i]$ and $\mathbf{x}\leq U_i$. Whenever $q$ exceeds
$b^U_i$ or falls below $b^L_i$, it switches to a new mode ($M_j$ or
$M_k$). When the timeout $U_i$ is reached, the transition to $M_p$ is
forced by the invariant. In both $S_i$ and $S_{i1}$, whenever the
automaton receives a synchronization signal \textbf{a?}, it must
immediately transit to $M_l$.

\subsection{PMC Models at Coarser Granularity}
\label{sec:pmc-models-at}

\pourreftex{
\subsubsection{Unoptimized Coarse Models}
}
\label{sec:simple}


We now give the translation scheme from \MTA{}s to TA at granularity
$g$ for $g>1$. This translation gives an abstraction of the original
\MTA, which consumes and produces coarse events. Like in the previous
section, we translate one PMC into two TA: the processing element
(PE) and the service model (SM).

\paragraph{Coarse Service Model (SM).}

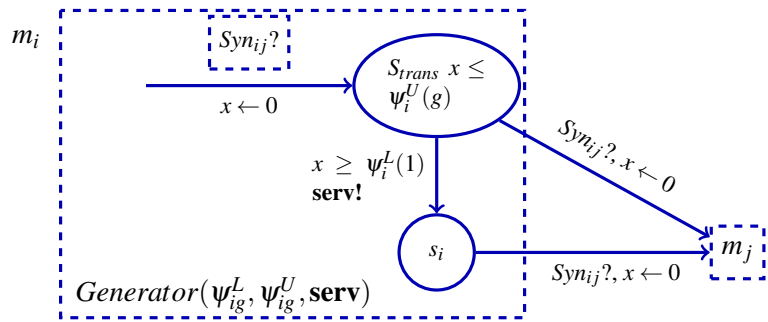
\begin{figure}[htbp]
  \centering
  
  \begin{tikzpicture}[node distance=4cm]
    \node[TAstate] (si) {$s_i$};
    \node[mode] (mj) [right of=si] {$m_j$};
    \node[TAstate,ellipse] (strans) [above=1cm of si,text width=1.3cm]
    {$S_{trans}$ $x\leq \psi_i^U(g)$};
    \node (before) [left of=strans] {};

    \path[TAtrans] (strans) -- (mj) node[midway,sloped,above] {$Syn_{ij}?$,
      $x\leftarrow 0$};
    \path[TAtrans] (strans) -- (si) node[midway,left, text width=1.5cm] 
    {$x\geq \psi_i^L(1)$ \textbf{serv!}};

    \path[TAtrans] (si) -- (mj) node[pos=.6,sloped,below]
    {$Syn_{ij}?$, $x\leftarrow 0$};
    \path[TAtrans] (before) -- (strans) node[midway,below]
    {$x\leftarrow 0$} node[midway,above=2mm,mode] {$Syn_{ij}?$};

    \node[label={below left:$m_i$}] (corner) at ($(before)+(-1,1cm)$) {};
    \coordinate (cornerD) at ($(si.south east)+(0.8cm,-0.5cm)$);
    \path[mode]  (corner) rectangle (cornerD);
    \node at ($(cornerD)+(-4,0.3)$)  
    {$Generator(\psi_{ig}^L, \psi_{ig}^U,\mathbf{serv})$};
  \end{tikzpicture}

  \caption{Simple coarse service model}
  \label{fig:simplecoarseservice}
\end{figure}

Changing the granularity for the SM is done by sampling the service
curves like we did 
in section~\ref{sec:sampling}.
It thus still uses one $Generator(\hat{\xi}^L_g$, $\hat{\xi}^U_g,
\text{\bf serv})$ of Figure~\ref{fig:generator} per mode $M_i$ but each
mode is refined into two states.
%
When a mode switch occurs, we do not know how many fine events have
been emitted since the last coarse \textbf{serv!} in previous mode
(but it has to be within the $[0, g)$). Arriving in the new mode
$M_i$, the SM has to wait for $k$ more fine events before emitting the
next coarse \textbf{serv!}, $k$ being in $(0, g]$. As shown in
Figure~\ref{fig:simplecoarseservice}, we add a state $S_{trans}$ to
capture this fact. The time the SM stays in this state is
non-deterministically chosen within $[\psi^L_i(1), \psi^U_i(g)]$; it
is measured by the clock \textbf{x}.

\paragraph{Coarse Processing Element (PE).}

The coarse translation of the PE is given in
Figure~\ref{fig:simpabspe} using the notations introduced in
Figure~\ref{fig:defstate}. The overall idea is the same as the fine
model (Figure~\ref{fig:finePE}), but the buffer fill level, here noted
$Q$, now counts coarse events instead of fine events. The state
$S_{i1}$ has been split into three states $S_{i1}$, $S_{inc}$ and
$S_{dec}$ whose meaning are given in the figure. This splitting is
necessary because when the condition between two modes depends on a
threshold $b^U_i$ or $b^L_i$ on the backlog $q$ at fine granularity,
if this threshold is not a multiple of $g$, then the actual transition
of the fine PMC would occur between two coarse events. We note $Y^L$
and $Y^U$ (resp. $H^L$ and $H^U$) the smallest and greatest numbers
of coarse events between which the fine threshold $b^U_i$
(resp. $b^L_i$) can be reached. The explanation and the values on
those constants are given Figure~\ref{fig:HY}.  We can therefore not
determine precisely when a transition due to a threshold \emph{does}
occur in the fine PE, but we have to ensure in the coarse PE, due to
the proof obligation~\ref{lemma:pmc}, that it \emph{can} occur at the
same time as it would have in the fine PE. 
The coarse PE leaves the state $S_{i1}$ when one of the transitions to
$M_j$ or $M_k$ \emph{can} occur, and the invariants on states
$S_{inc}$ and $S_{dec}$ say when the transition \emph{must} occur.
The management of synchronization events (\textbf{a?}) and timeout
($\mathbf{x}=U_i$) is the same as in the fine model. For clarity, it
is not drawn on the TA but described above.


\begin{figure}[htbp]
  {\centering

  \begin{tikzpicture}[
    font=\footnotesize]
    \node[PMCstate] (Si) [
    label=above left:{$[-\infty,\infty]$},
    ] {$x\leq L_i$};
    \node (nameSi) at (Si) [above=1mm] {$S_i$}; 

    \coordinate[left of=Si,node distance=3cm] (Mr);
    \node[dashed, draw, draw=blue!70!black, very thick, minimum
    size=1.5cm] (Mrbis) at (Mr) {};
    \path ($(Mr)+(-.4,-.4)$) node {$M_r$};
    \path ($(Mr)+(.35,.1)$) node[above=1pt] {$Syn_{ri}!$};
    \path[TAtrans,dashed] (Mr) -- (Si) node[pos=.7,below]
    {$x\leftarrow 0$};
 
    \node[PMCstate, right of=Si,node distance=5cm] (Si1) [
    label=right:{$[H^U+1,Y^L-1]$},
    ] {$x\leq U_i$};
    \node (nameSi1) at (Si1) [above=1mm] {$S_{i1}$}; 

    \node[PMCstate, above of=Si1,node distance=2.5cm] (Sinc) [
    label=above:{$[Y^L,Y^U]$},
    ] {$x\leq U_i$};
    \node (nameSinc) at (Sinc) [above=1mm] {$S_{inc}$}; 

    \node[PMCstate, below of=Si1,node distance=2.5cm] (Sdec) [
    label=below:{$[H^L,H^U]$},
    ] {$x\leq U_i$};
    \node (nameSdec) at (Sdec) [above=1mm] {$S_{dec}$}; 
    
    \node[mode, right of=Sinc,node distance=4.5cm] (Mj) {$M_j$};
    \node[mode, right of=Sdec,node distance=4.5cm] (Mk) {$M_k$};

    \path[TAtrans] (Si) -- (Si1) node[midway,above] {$x=L_i\wedge
      H^U< Q< Y^L$};
    \path[TAtrans] (Si) .. controls ++(1,1.5) .. (Sinc) 
    node[pos=.8,above,sloped] {$x=L_i\wedge Y^L\leq Q \leq Y^U$};
    \path[TAtrans] (Si) .. controls ++(1,-1.5) .. (Sdec) 
    node[pos=.8,above,sloped] {$x=L_i\wedge H^L\leq Q \leq H^U$};

    \path[TAtrans] (Sinc) -- (Mj) node[midway,above] {$Syn_{ij}!$};
    \path[TAtrans] (Sdec) -- (Mk) node[midway,above] {$Syn_{ik}!$};

    \path[TAtrans] (Si1) ..controls++(.5,1.3) .. (Sinc) 
    node[midway,right] {$Q=Y^L-1$};
    \path[TAtrans] (Sinc) ..controls++(-.5,-1.3) .. (Si1) 
    node[midway,left] {$Q=Y^L$};

    \path[TAtrans] (Si1) ..controls++(.5,-1.3) .. (Sdec) 
    node[midway,right] {$Q=H^U+1$};
    \path[TAtrans] (Sdec) ..controls++(-.5,1.3) .. (Si1)
    node[midway,left] {$Q=H^U$};

    \path[TAtrans] (Si) |- ++(0,3.8) -| (Mj) node[pos=.1,below] 
    {$x=L_i\wedge Q>Y^U$ $Syn_{ij}!$};
    \path[TAtrans] (Si) |- ++(0,-3.8) -| (Mk) node[pos=.1,above] 
    {$x=L_i\wedge Q<H^L$ $Syn_{ik}!$};
  \end{tikzpicture}

  }
  \vspace*{0.5\baselineskip}
  
  {\sl\underline{Coarse thresholds:}
    $Y^L = \big\lfloor (b_i^U+1)/g\big\rfloor, \;
    Y^U = \big\lceil (b_i^U+1)/g\big\rceil,\;  
    %
    H^L = \big\lfloor (b_i^L-1)/g\big\rfloor, \;
    H^U = \big\lceil (b_i^L-1)/g\big\rceil
    $\\[0.5\baselineskip]
    \underline{Explanation on those values:} let $N_r$ (resp. $N_s$)
    denotes the total number of \emph{coarse} events \textbf{req?}
    (resp. \textbf{serv?})  received since the system started. Let
    $n_r$ (resp.  $n_s$) denotes the corresponding total number of
    \emph{fine} events.  $Q$ (resp. $q$) denotes the \emph{coarse}
    (resp. \emph{fine}) buffer fill level. At any time, we have the
    following constraints:
    $g N_r \leq n_r < g(N_r+1) \;\text{ and }\; g N_s \leq n_s <
    g(N_s+1)$. \\
    Since the fine and coarse buffer fill levels $q$ and $Q$ satisfy
    $q=n_r-n_s$ and $Q=N_r-N_s$ respectively, we deduce that $g(Q-1) <
    q < g(Q+1)$. This implies that 
    $\lfloor q/g\rfloor \leq Q \leq \lceil q/g \rceil$.  When $q$ is
    replaced with $b^U_i+1$ in the former inequation, it provides
    lower and upper bounds $[Y^L, Y^U]$ on the value of $Q$.  We can
    similarly derived the bounds $[H^L, H^U]$.  \\[0.5\baselineskip]}
  \label{fig:HY}
  %
  %
  %
  {\sl
    \underline{Meaning of the states:}
    $\mathbf{S_i}$:  same as $S_i$ in the fine PE.
    $\mathbf{S_{i1}}$: the coarse backlog corresponds to a fine buffer
    fill level between $b_i^L$ and $b_i^U$.
    $\mathbf{S_{inc}}$ / $\mathbf{S_{dec}}$: in the fine PE, the
    buffer fill level may have reached $b_i^U$ / $b_i^L$. \\[0.5\baselineskip]
    %
    \underline{Other transitions:} 4 transitions labelled
    (\textbf{a?},\textbf{Syn$_{il}$!}) from $S_i$, $S_{inc}$, $S_{i1}$,
    $S_{dec}$ to the mode $M_l$ and 3 transitions labelled
    ($\mathbf{x}=U_i$,\textbf{Syn$_{ip}$!}) from  $S_{inc}$, $S_{i1}$,
    $S_{dec}$ to the mode $M_p$.
  }
  \caption{Simple coarse PE model}
\label{fig:simpabspe}
\end{figure}

\paragraph{Validation.}

Due to the proof obligation~\ref{lemma:pmc}, we now have to prove
that the proposed simple coarse PMC provides a correct abstraction
of the fine PMC.

\noindent\textit{Proof:}
let $\hat{\tau}=(\hat{t}_0)\hat{t}_1\hat{t}_2...$ be any fine input
stream to the fine PMC $\model_0$, and
$\breve{\tau}=(\breve{t}_0)\breve{t}_1...$ is the corresponding output
stream. Firstly, we construct a coarse input stream
$\hat{\top}=(\hat{T}_0)\hat{T}_1\hat{T}_2...$ with
$\hat{T}_j=\hat{t}_{gj}$ for $j\geq 0$. It is clear that $\hat{\top}$ can
be refined to $\hat{\tau}$, while $\hat{\top}$ can be an input to the
coarse PMC $\model_C$ due to:

{\small
\centering
\(\forall j\geq k\geq 0,
  \hat{\xi}_{g}^L(j-k) = \hat{\xi}^L(g(j-k)) \leq   \hat{t}_{gj}-\hat{t}_{gk}
  \leq \hat{\xi}^U(g(j-k)) = \hat{\xi}_g^U(j-k)
\)

}

%
Then we show how to construct a coarse output stream $\breve{\top}$ from
$\model_C$ corresponding to the input stream $\hat{T}$ and which is an
abstraction of $\breve{\tau}$: given the behavior of $\breve{\tau}$,
we build, by induction, the behavior $\breve{\top}$ where mode
switches occur at the same time in the fine and the coarse PMCs.

Firstly, when the system starts, both PMCs can enter the starting mode
at the same time.
Suppose now that the coarse output stream $\breve{\top}$ has
been constructed until the $(A-1)$-th event, for some $A>0$, to be
$(\breve{T}_0)\breve{T}_1...\breve{T}_{A-1}$ with
$\breve{T}_j = \breve{t}_{gj}$ for all $j\in[0,A-1]$; and that both
$\model_0$ and $\model_C$ entered the mode $M_i$ at the same instant.

Let the fine PMC $\model_0$ leaves $M_i$ when the ($gB+n$)-th fine
event \textbf{produce!} is processed, with $n<g$. Two cases can then happen:

{\noindent\bf Case 1:} $B=A-1$. From the service model shown in
Figure~\ref{fig:simplecoarseservice}, the time to generate the next
coarse event \textbf{serv?} is lower and upper bounded by $\psi^L_i(1)$
and $\psi^U_i(g)$ respectively. It covers all the possibility of what
can happen in the fine service model. Hence, it is possible that the
coarse PMC leaves $M_i$ at the same time as the fine one.

{\noindent\bf Case 2:} $B>A-1$. In mode $M_i$, we continue to
construct $\breve{\top}$ to be
$...\breve{T}_{A-1}\breve{T}_A...\breve{T}_B$, with $\breve{T}_j =
\breve{t}_{gj}$ for $j\in[A,B]$. Similarly to the {\em Case 1} above,
we can show that $\breve{T}_A$ can be an output from $\model_C$. We
can also show that $\breve{T}_{A+1}...\breve{T}_B$ can be an output
from $\model_c$, due to
$\psi^L_i(g(j-A))  \leq \breve{t}_{gj} - \breve{t}_{gA} \leq  \psi^U_i(g(j-A))$
and $\psi^L_{gi}(j-A) = \psi^L_i(g(j-A))$,
$\psi^U_i(g(j-A))=\psi^U_{gi}(j-A)$. The coarse PMC is able to leave
$M_i$ at the same time as the fine PMC. Indeed, as shown in
Figure~\ref{fig:simpabspe}, $\model_C$ may transit out from $S_{inc}$
(or $S_{dec}$) anytime before $Q$ increases to $Y^U+1$ (or falls below
$H^L-1$). Hence, it is possible that $\model_C$ transits out at the
same time as $\model_0$ for the case of those transitions. In other
cases of transiting out, the time to transit out is same for
$\model_C$ and $\model_0$, which is equal to $L_i$, $U_i$ or the time
receiving the synchronization signal \textbf{a?}.

It is clear that the constructed coarse output stream $\breve{\top}$ is
an abstraction of the fine one $\breve{\tau}$. Therefore, the proof
obligation~\ref{lemma:pmc} is validated.
%



\paragraph{Optimization of the coarse PMC model.}



This unoptimized model introduces non-determinism 
when mode changes occur. For example, if one knows that a mode change
occurs when $q=12$, with a granularity of $g=5$, one can not capture
in the coarse model the exact instant where $q$ becomes equal to 12,
but only the arrival of the second and third coarse events ($T_2$ and
$T_3$), which correspond to $q=10$ and $q=15$.
By reusing the information we have on the service and arrival curves
at the fine granularity, we can get a better estimation than $[T_2,
T_3]$ for the time at which the mode change occur. In the example, one
can get a lower bound $x$ for the time needed to get 2 more events in
the buffer: $\hat{\xi}^U$ says how fast the events can arrive, and
$\psi^L$ says how fast the PMC processes them. We can compute the
minimal time for which the difference between the number of events
received and the number of events processed is 2.
%
Similarly, we can get an interval $[i_s, i_e]$ on the time between the
mode change and the next coarse \textbf{serv?} event. In the TA model,
this is implemented by forcing the automaton to remain in the old mode
for $x$ time units after receiving the second event, and to enable the
transition for the first \textbf{serv?}  only when the time spent in
the mode is in $[i_s, i_e]$.

We implemented two variants of this optimization in our prototype. The
first introduces a counter $\omega$ and the corresponding coarse PMC
is called $\model_C$-$\omega$, and the second does all the complex
computations on best and worst cases using this counter, and the
corresponding coarse PMC is called $\model_C$-opt. These
optimizations considerably increase the precision of the
analysis. They are detailed in~\cite{verimag-TR-2009-10},
but omitted here by lack of space.


\section{Experimental Validation}
\label{sec:exp}
\label{sec:example}

\paragraph{Tools.}
To conduct experimentations on our framework, we applied the TA
modeling and verification tool UPPAAL CORA \cite{cora}, 
to
model TA and to analyze the output arrival curves by model
checking. The generators and observers are written once for all, but for
the PMC, the \MTA translation into TA has still to be written by
hand. Running the analysis at multiple granularities and the algorithm
to combine the obtained curves into the tightest one is fully
automated.

\pourreftex{
\subsection{An Example of PMC}
}

\paragraph{An Example of PMC.}

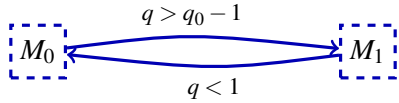
\begin{figure}[tbp]
  \centering

  \begin{tikzpicture}
    \node[mode] (mo) {$M_0$};
   \node[mode] (ml) at (mo) [right=4cm] {$M_1$};

    \path[TAtrans] (mo) .. controls ++(2,0.25) .. (ml)
    node[midway,above]
    {$q>q_0-1$};

    \path[TAtrans] (ml) .. controls ++(-2,-0.25) .. (mo)
    node[midway,below]
    {$q<1$};
  \end{tikzpicture}

  \caption{\MTA model of the example PMC}
  \label{fig:expdesp}
\end{figure}

We illustrate the approach with the fine and coarse models of a simple
PMC illustrating a common behavior. It runs at two modes:
``sleep'' and ``run''. It only processes events and consumes energy in
the ``run'' mode and switches to the power-saving ``sleep'' mode when
its buffer is empty. To avoid switching back and forth between ``run''
and ``sleep'', the PMC waits until its buffer fill level $q$ reaches a
threshold $q_0$ before transiting from ``sleep'' to ``run''. Initially
the input buffer is empty (i.e. $q = 0$). Figure~\ref{fig:expdesp}
shows the model of this example PE in terms of \MTA.

\pourreftex{
\subsection{Result of the Analysis for the Example}
}

\begin{figure}[tbp]
  \centering
  \includegraphics[scale=0.6]{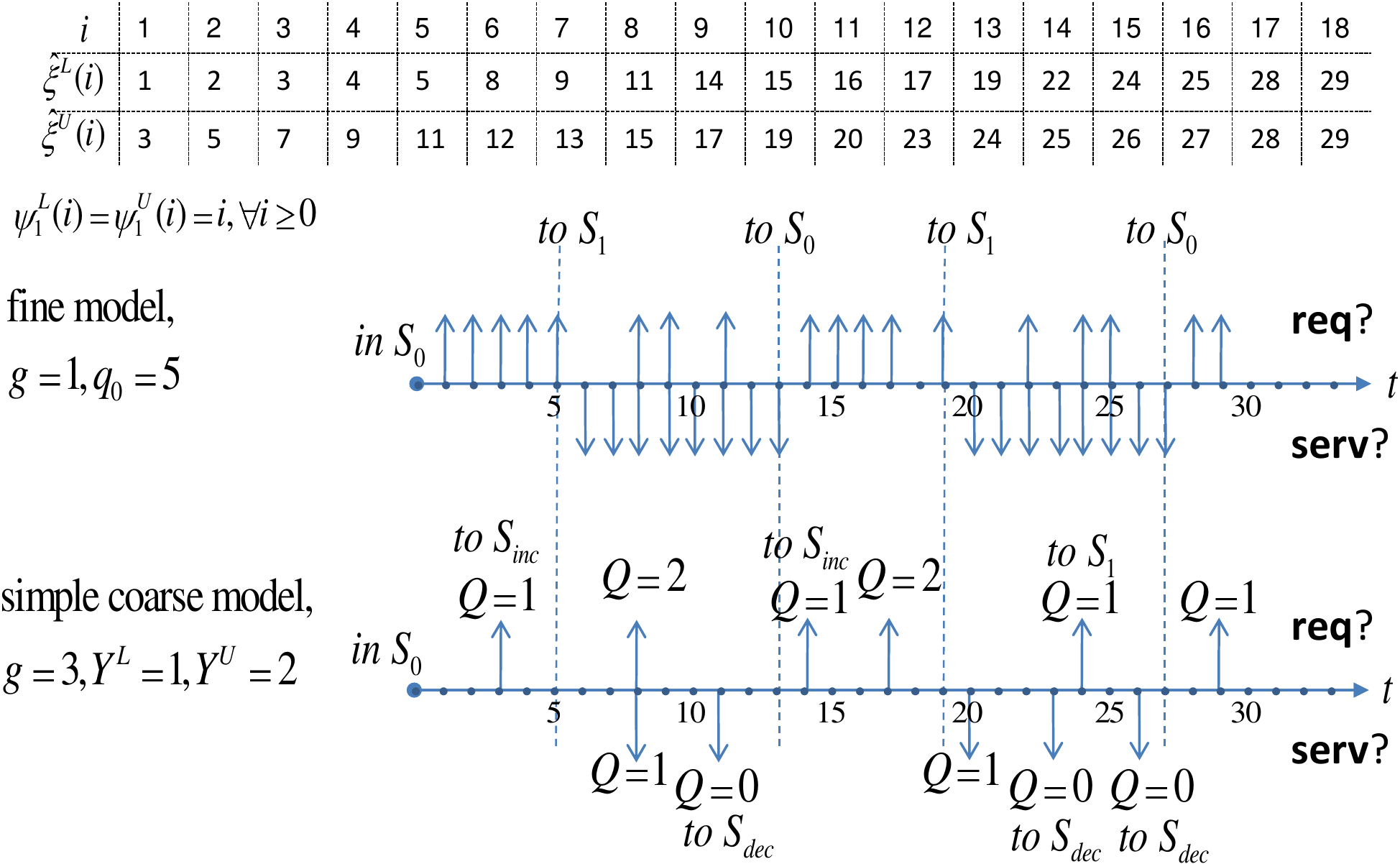}
  \caption{An illustrated flow of the example PMC}
  \label{fig:expflow}
\end{figure}

\paragraph{Result of the Analysis for the Example.}

We show the experiments conducted 
on the previous example where the threshold $q_0$ is set to 5
(i.e. the PMC starts to run when the number of fine events in the
buffer reaches 5). 
%
We first apply the translation from the \MTA to TA presented above to
get the coarse and fine-grain model. Notice that, since the \MTA
doesn't exhibit all the cases of mode-changes, the translation can
indeed be optimized manually (see
~\cite{verimag-TR-2009-10}).
In Figure~\ref{fig:expflow}, we take an example input stream and show
its execution in the fine and coarse PMC models at granularity 3. It
can be observed that the coarse PE always switches from $M_0$ to $M_1$
when it stays in $S_{inc}$ and switches from $M_1$ to $M_0$ when it
stays in $S_{dec}$. This illustrates how the non-determinism is
introduced in the coarse model.

{\small
\begin{table*}[htbp]
  \begin{center}
    \def\m{\model}
\begin{tabular}{|c|c||c|c|c||c|c|c||c|c|c|}
\hline
output                & \multicolumn{10}{|c|}{analysis time at granularity $g$ [sec]}                                                                            \\
\cline{2-11}
arrival               & $g=1$    & \multicolumn{3}{|c||}{$g=2$}              & \multicolumn{3}{|c||}{$g=3$}            & \multicolumn{3}{|c||}{$g=4$}            \\
\cline{2-11}
curves                & $\m_0$   & $\m_C$   & $\m_C$-$\omega$ & $\m_C$-opt & $\m_C$ & $\m_C$-$\omega$ & $\m_C$-opt & $\m_C$ & $\m_C$-$\omega$ & $\m_C$-opt \\
\hline
$\breve{\xi}^L_g(k)$  & 23674.4  & 56.1       & 62.9            & 223.6      & 7.90     & 12.9            & 189.7      & 0.76     & 1.21            & 28.6       \\
\hline
$\breve{\xi'}^U_g(k)$ & 411515.3 & 746.2      & 583.2           & 4217.7     & 68.6     & 120.9           & 1465.4     & 7.43     & 12.5            & 269.1      \\
\hline\hline
distance              & /        & 4.17       & 2.46            & 0.63       & 5.50     & 2.88            & 0.63       & 9.08     & 5.08            & 1.25       \\
\hline
\end{tabular}
\end{center}
    \def\mean{\text{\it mean}}
\caption{Time to compute fine and coarse output
    arrival curves ($\breve{\xi}^L_g$, $\breve{\xi'}^U_g$),
    and the distance between coarse and fine curves, with 
    $ distance = \mean( \mean (\breve{\xi}^L(gk)-\breve{\xi}^L_g(k)),
    \mean(\breve{\xi}^U_g(k)-\breve{\xi}^U(gk)) )
    $
    where $\mean$ is  the average of all the elements for $1\leq
    k \leq \lfloor 24/g \rfloor$.}
\label{tab:time}
\end{table*}
}

We now comment the results of coarse output arrival curves when
running the analysis  at
multiple granularities $g=2,3,4$.  First the input arrival curves
$\hat{\xi}$ and service curves $\psi$ are sampled by 
taking the points of the curves having an abscissa multiple of $g$ (an
illustration is given in appendix~\ref{sec:arrival-service}).
Then, using the UPPAAL CORA tool, we analyze the minimum cost to reach
the \texttt{Stop} state of the observer model, which gives the lower output
curve $\breve{\xi}^L_g(k)$ and use the variant of the observer shown
in Figure~\ref{fig:observer}(b) to compute $\breve{\alpha}^U$, which
gives $\breve{\xi}^U_g$ after pseudo-inversion.  

For granularity $g=4$, we compare the output curves
($\breve{\xi}^L_4$, $\breve{\xi}^U_4$) using the unoptimized PMC
$\model_C$, ($\breve{\xi}^L_4$-$\omega$, $\breve{\xi}^U_4$-$\omega$)
using the model with $\omega$, $\model_C$-$\omega$ and
($\breve{\xi}^L_4$-opt, $\breve{\xi}^U_4$-opt) using the optimized
PMC $\model_c$-opt, as shown in
Appendix~\ref{sec:output-curves-comparison}. It
can be observed that $\model_C$-$\omega$ helps to obtain tighter
coarse output curves than $\model_C$, which are further improved by
$\model_C$-opt.

When analyzing the coarse output curves at each granularity,
$g=2,3,4$, obtained from the optimized coarse PMC $\model_C$-opt,
($\breve{\xi}^L_g$, $\breve{\xi}^U_g$), it can be observed that
$\breve{\xi}^L_g(k)$ provides a lower bound on $\breve{\xi}^L(gk)$ and
$\breve{\xi}^U_g(k)$ provides an upper bound on $\breve{\xi}^U(gk)$;
where ($\breve{\xi}^L$, $\breve{\xi}^U$) is the fine output curves
computed from the fine PMC $\model_0$ 
(see Appendix~\ref{sec:output-curves-comparison}).

Table~\ref{tab:time} summarizes the analysis at multiple granularities
by showing the total time to compute the fine and coarse output curves
($\breve{\xi}^L_g(k)$, $\breve{\xi}^U_g(k)$) for $g=1,2,3,4$ and
$k\leq \lfloor 24/g \rfloor$. It also shows the $distance$ measured
between coarse and fine curves.  As expected, the three models
$\model_C$, $\model_C$-$\omega$ and $\model_C$-opt allow a trade-off
between performance and accuracy, $\model_C$ being the fastest and less
precise, and $\model_C$-opt the slowest and most precise. The
granularity $g$ allows another trade-off: the coarsest models are the
least precise ones and the fastest to analyze.

Finally, we experiment that applying the causality closure
\cite{causality-TACAS10}, (quadratic algorithm), to the resulting
curves give information that neither of the analysis would have given
alone. For example, running the analysis with $q_0=21$ at
granularities $g=9$ and $g=10$, we get $\breve{\xi}^U_{10}(2)=111$ and
$\breve{\xi}^U_9(2)=108$, which trivially implies $\breve{\xi}^U(10)
\leq 108$. Combining the curves and using the information provided by
$\breve{\xi'}^L$, we get the value $\breve{\xi}^U(10) = 102$. The
complete curve is in appendix~\ref{sec:combine-and-refine}.



\section{Conclusion}
\label{sec:conclusion}

In this paper, we have proposed a novel framework of granularity-based
interfacing between RTC and TA performance models, which complements
the existing work and reduces the complexity of analyzing a
state-based component modeled by TA. We have illustrated the approach
with an example which shows how the model of a component is abstracted
to work with an event stream at coarse granularity. We did experiments
that show how the abstraction is validated: they confirm that the
precision of the results depends on a tradeoff with the analysis time.
Indeed, the timing results show that the time to analyze the coarse
models reduces at least $99\%$ of that for the fine
models. Furthermore, using the results from multiple runs of analysis
at different granularities, we also demonstrate how to obtain bounds
on the arrival patterns of the fine output stream with a reasonable
loss of precision.

In future works, 
We aim at further characterizing the class of \MTA: from a theoretical
point of view, we may compare it to some existing class, such as event
count automata \cite{CPT05}; and from a practical point of view, we
will go on checking if the expressivity of \MTA is enough to model
more complex power-managed components.  We also plan to adapt the
granularity changes to other state-based models, namely the ones in
\cite{verimag-TR-2009-14}.
On the other hand, a more challenging perspective would be to work on
new state-based abstraction techniques for analyzing the time
\emph{and} energy consumption of a component.

\bibliographystyle{eptcs}
\bibliography{gran-paper}
\FloatBarrier
\appendix












\section{Arrival and Service Curves}

\subsection{Input Curves}
\label{sec:arrival-service}

\begin{figure}[H]
  \centering
  \includegraphics[scale=0.6]{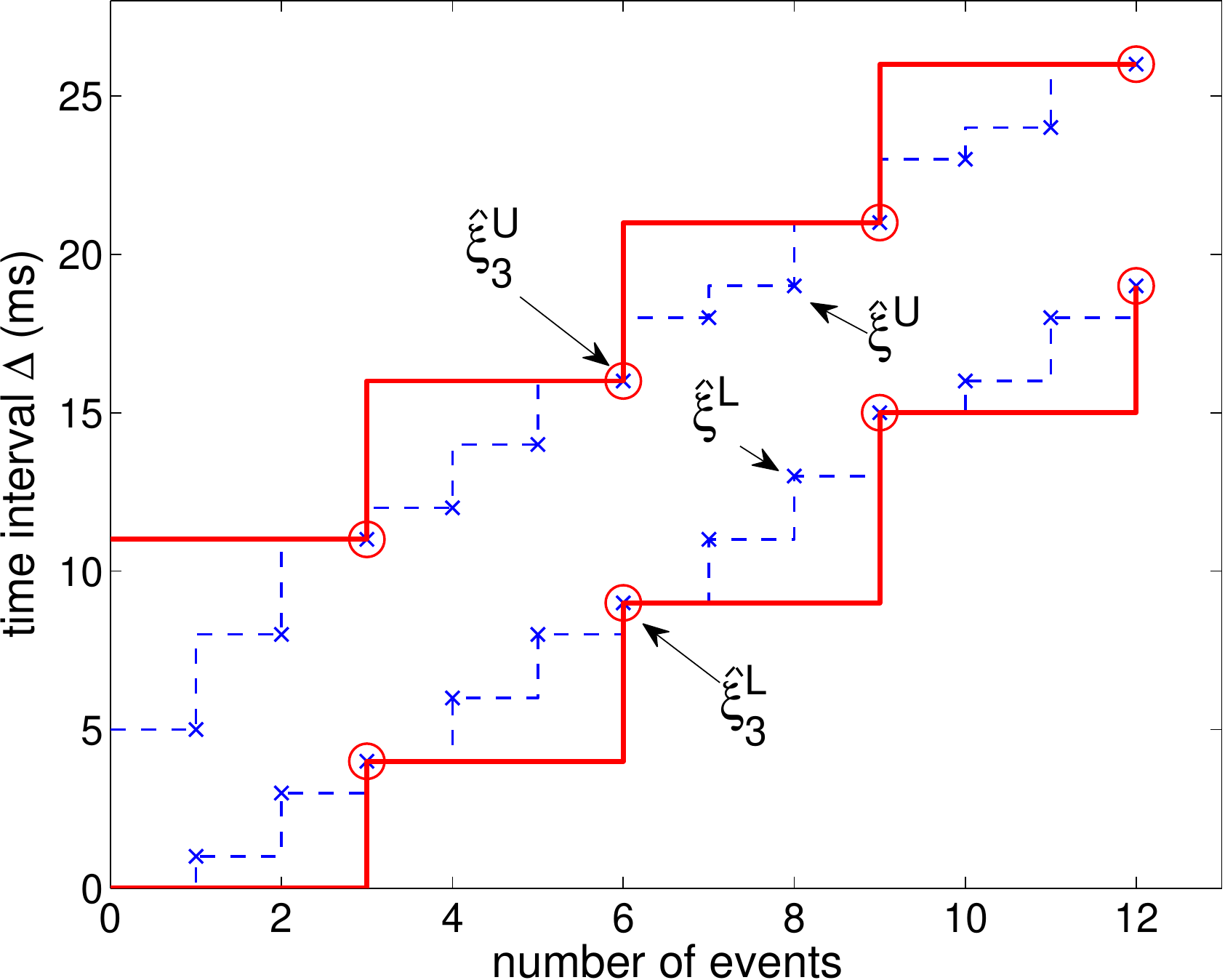}
  \caption{Example of fine and coarse input arrival curves
    $\hat{\xi}^{L/U}$ and $\hat{\xi}^{L/U}_3$}
  \label{fig:arrival}
\end{figure}

\begin{figure}[H]
  \centering 
  \includegraphics[scale=0.6]{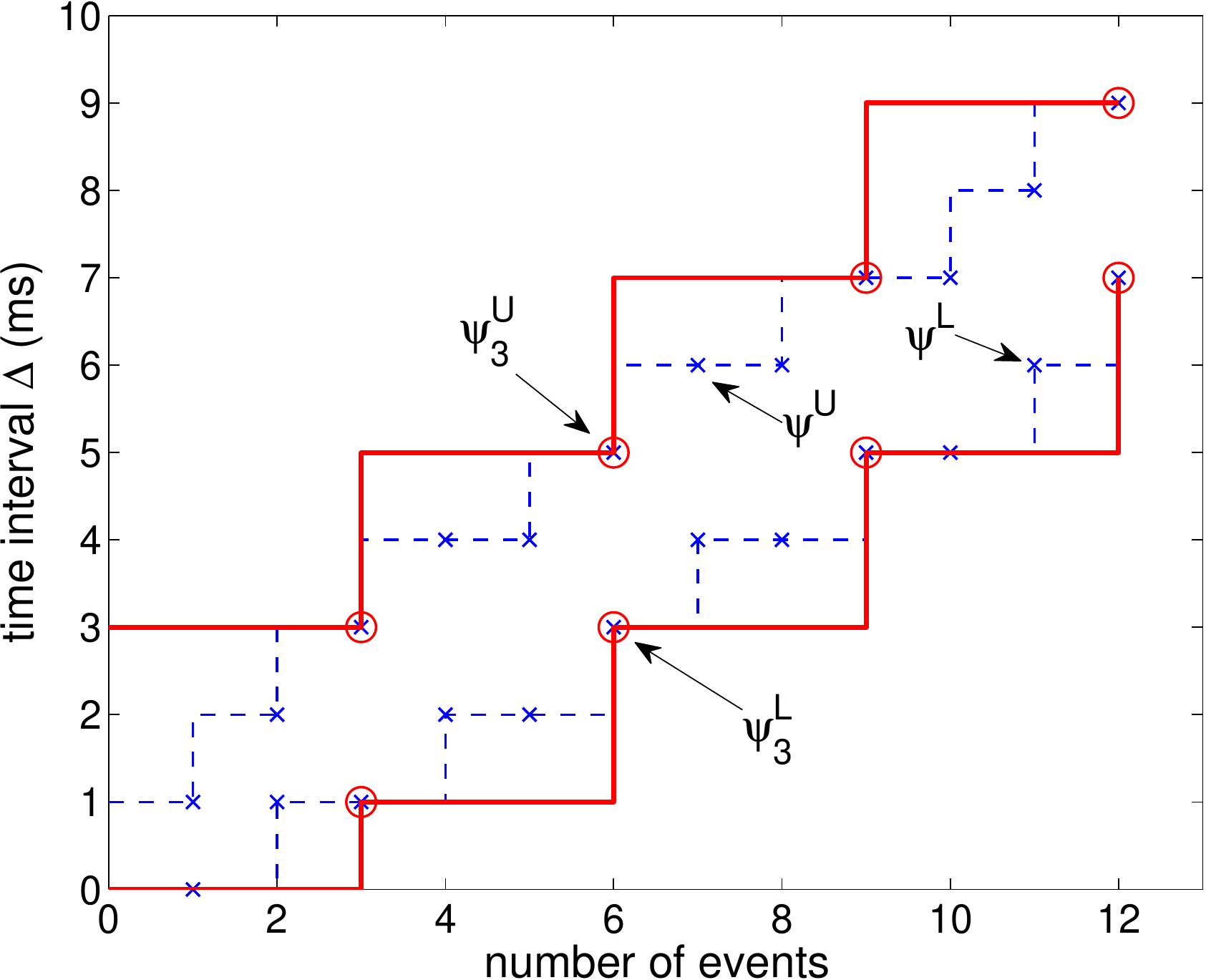}
  \caption{Example of fine and coarse service curves $\psi^{L/U}$
    and $\psi^{L/U}_3$}
  \label{fig:service}
\end{figure}

\subsection{Output Curves}
\label{sec:output-curves-comparison}

\begin{figure}[H]
  \centering 
  \includegraphics[scale=0.6]{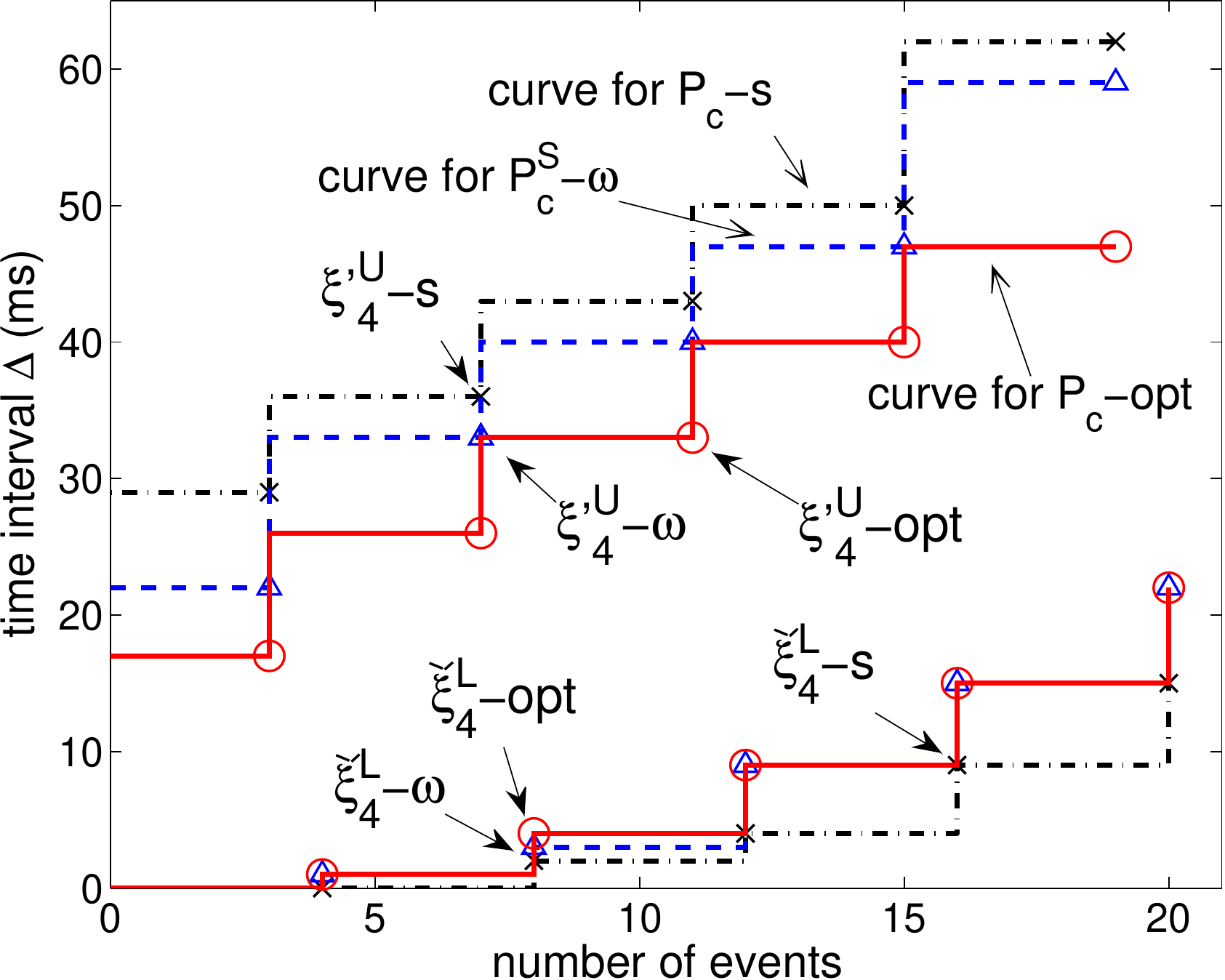}
  \caption{Comparison of output arrival curves at $g=4$ using
    simple coarse PMC $\model_c$, coarse PMC with $\omega$
    $\model_c$-$\omega$ and optimized PMC $\model_c$-opt}
  \label{fig:compareopt}
\end{figure}

\begin{figure}[H]
  \centering
  \includegraphics[scale=0.6]{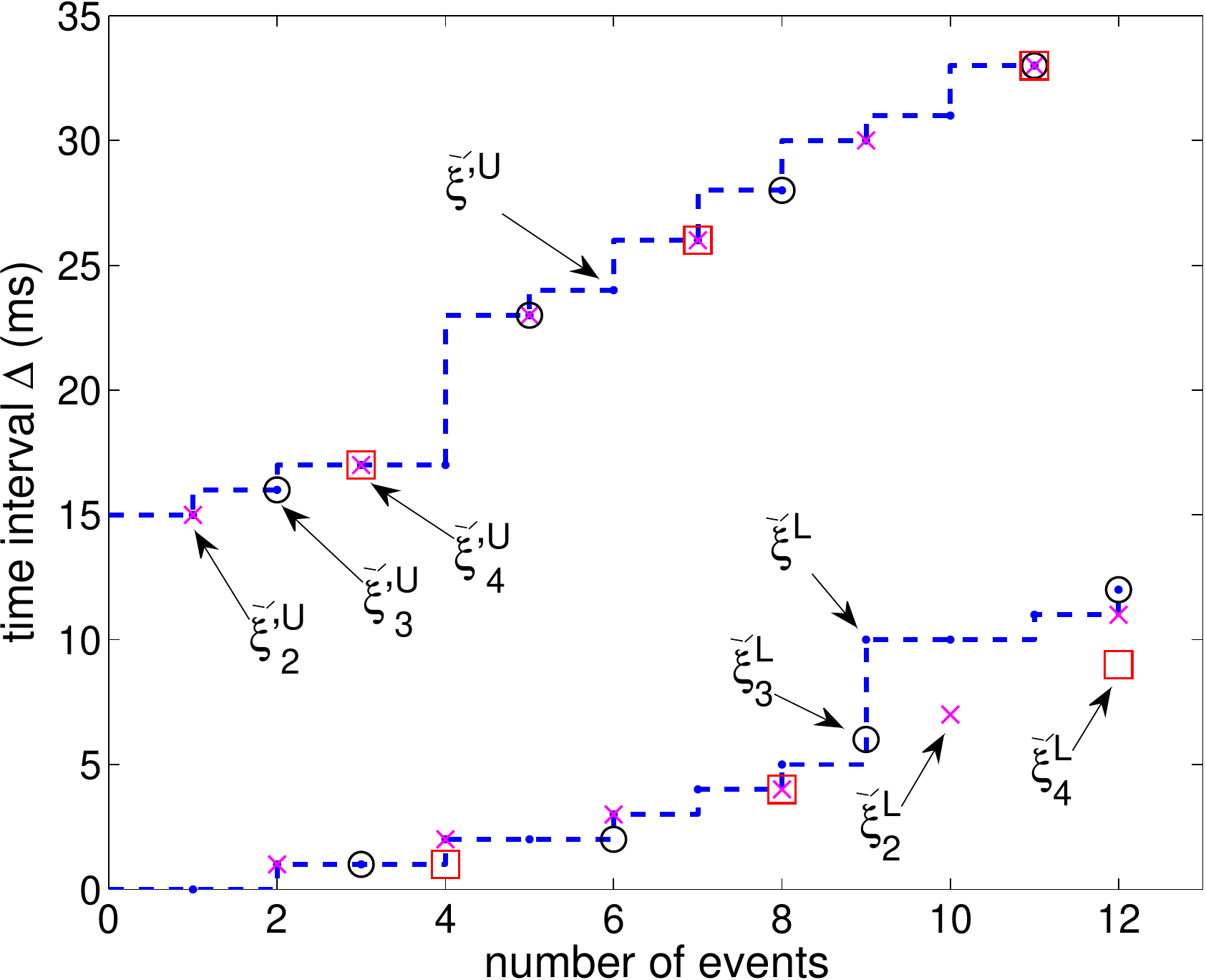}
  \caption{Comparison between fine and coarse output arrival
    curves computed from fine and optimized coarse PMC models
    respectively}
  \label{fig:output}
\end{figure}

\subsection{Combined Output Curves}
\label{sec:combine-and-refine}

\begin{figure}[H]
  \centering
  \includegraphics[scale=0.6]{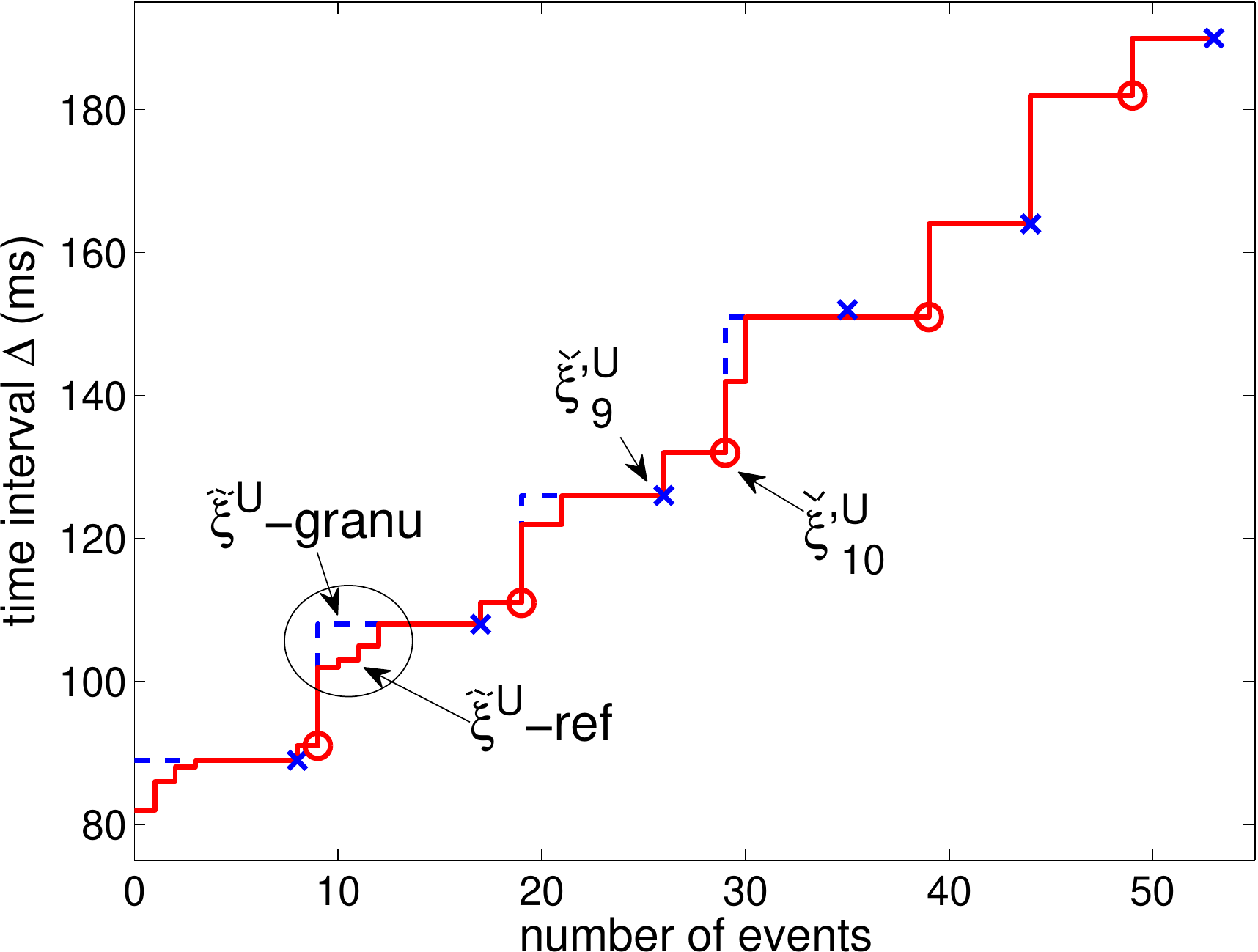}
  \caption{Computed fine upper output arrival curves
    $\tilde{\xi}^U$-granu (with simple scheme) and
    $\tilde{\xi}^U$-ref (with mathematical refinement algorithm)}
  \label{fig:refine}
\end{figure}

\end{document}